\documentclass[aip,jcp,reprint,amsmath,amssymb]{revtex4-2}

\usepackage{graphicx,siunitx,xcolor,hyperref}
\usepackage[version=4]{mhchem}
\hypersetup{
colorlinks,
linkcolor={blue!50!black},
citecolor={blue!50!black},
urlcolor={blue!80!black}}

\begin{document}
\title{Unveiling plastic rearrangements in silica glass through persistent homology}
\author{Fuya Katsuyama}
\email{f.katsuyama71@sanken.osaka-u.ac.jp}
\author{Kumpei Shiraishi}
\author{Emi Minamitani}
\affiliation{SANKEN, The University of Osaka, Ibaraki, Osaka 567-0047, Japan}

\begin{abstract}
Plastic deformation of amorphous solids proceeds through localised atomic rearrangements.
While much effort has been devoted to identifying where such rearrangements occur, how the local structure changes during a rearrangement is less well understood.
In silica glass, a rearrangement can involve both distortion within \ce{SiO4} tetrahedra and deformation of the network connecting them, but the contributions of these two modes have rarely been quantified.
Here, we characterise plastic rearrangements in silica glass under athermal quasistatic shear using persistent homology, which extracts rings from atomic configurations, and link the changes in the persistence diagrams to individual rings in real space.
We classify rearrangement events into two types according to bond-coordination changes.
By performing inverse analysis, we reveal that \ce{Si}-centred triangles within tetrahedra, hinge triangles spanning two corner-sharing tetrahedra, and the other (non-hinge) \ce{O3} triangles are the dominant sources of topological changes upon rearrangements.
The birth-time changes of \ce{Si}-centred and hinge triangles closely follow the change in the angle opposite the longest edge, whereas that of non-hinge triangles also reflects changes in inter-tetrahedral \ce{O}--\ce{O} distances.
Further, we clarify that triangles with large birth-time changes are localised in the rearrangement region.
Comparing the two types of triangles, we find that the ratio of their birth-time changes is larger for coordination-changing events than for coordination-preserving events, indicating that such events involve relatively more intra-tetrahedral deformation compared with network deformation.
These results demonstrate that persistent homology separates intra-tetrahedral and network deformation and quantifies each as a continuous quantity within a single framework.
\end{abstract}
\maketitle

\section{Introduction}
The plastic deformation of amorphous solids has been a long-standing problem in condensed matter physics~\cite{Falk_2011}.
In crystals, plasticity is carried by well-defined structural defects such as dislocations, whose motion governs the mechanical response~\cite{Hull2011}.
By contrast, amorphous solids lack long-range order, and no such defects can be identified a priori from the structure.
In amorphous solids, plastic deformation has instead been attributed to collective rearrangements of localised groups of particles~\cite{Spaepen1977,Argon1979,Falk1998}.
These rearrangements can trigger one another to form avalanches~\cite{Maloney_Lemaitre_2004,MaloneyLemaitre2006,PhysRevE.82.031301}, which may organise into shear bands~\cite{Dasgupta2012,Ozawa_2018} and eventually lead to macroscopic yielding~\cite{Falk_2011}.
In the quasistatic limit, individual rearrangements are separated by intervals of elastic loading and can therefore be isolated as discrete events~\cite{Maloney_Lemaitre_2004,MaloneyLemaitre2006}.
Considerable effort has been devoted to identifying where these rearrangements occur, using indicators such as locally favoured structures~\cite{Ding_2014}, low-frequency vibrational modes~\cite{Tanguy_2010,Manning2011,Rottler_2014,Ding_2014,Tong2019Revealing}, machine-learning techniques~\cite{Cubuk2015,Cubuk2017}, and local yield stress~\cite{Patinet_2016} (see Ref.~\onlinecite{Richard2020Predicting} for a systematic comparison).
In contrast, how the local structure itself changes during a rearrangement is less well understood.
Because no well-defined elementary process such as dislocation glide exists in amorphous solids, it is not obvious which structural motifs deform, and by how much, when a rearrangement occurs.
Characterising these structural changes is therefore essential for understanding the mechanical response of amorphous solids.

Much of this understanding has been obtained for glasses with non-directional interactions, such as Lennard-Jones mixtures and metallic glasses.
Silica glass (amorphous \ce{SiO2}), a prototypical network-forming glass, instead consists of corner-sharing \ce{SiO4} tetrahedra connected by directional covalent bonds~\cite{Zachariasen1932,Mozzi1969,Mazurin1983}.
Cascades of plastic rearrangements similar to those in non-network glasses have also been reported in silica glass~\cite{Pingua2025}.
In this covalent network, however, plastic rearrangements can involve not only deformation of the network connecting the tetrahedra but also changes within the tetrahedra themselves, including breaking and reforming of \ce{Si}--\ce{O} bonds, and rearrangements accompanied by such bond switching have been reported to produce larger energy drops~\cite{Koziatek2015,Bonfanti2019}.
Plastic deformation of silica glass can therefore proceed through two distinct modes, namely distortion within the tetrahedra and deformation of the network connecting them.
However, how much each mode contributes to a given rearrangement has rarely been quantified directly from atomic configurations~\cite{Mantisi2012}.

Quantifying these two modes requires a structural descriptor that resolves both the geometry within individual tetrahedra and that of the network connecting them.
Ring statistics have commonly been used to characterise the topology of network-forming glasses~\cite{KING1967,Guttman1990}.
However, rings are defined on the bond network, so ring statistics are insensitive to structural features not encoded in the bond connectivity, such as continuous distortions that leave all bonds intact~\cite{LeRoux2010}.
Moreover, the rings obtained depend on the choice of the bond cutoff distance and of the ring definition~\cite{Franzblau1991,LeRoux2010}.
A descriptor that does not rely on such choices and captures structural changes as continuous quantities is therefore desirable.

Persistent homology (PH), a method of topological data analysis, offers such a descriptor~\cite{Edelsbrunner2000,Zomorodian2005,Obayashi2022}.
PH extracts topological features such as rings from atomic configurations over multiple length scales without relying on a definition of bonds or a specific cutoff, and records each ring together with the length scales at which it appears and disappears.
PH has been applied to characterise the structures of amorphous materials, including silica glass, and has revealed features such as medium-range order~\cite{Nakamura2015,Hiraoka2016,Onodera2020,Soerensen2022,Minamitani2023,Firooz2024,Shiga2024,Minamitani2025,Yamazaki2026}.
PH has also been used to track the evolution of topological features under external stress or deformation~\cite{Shirai2019,Pan2024,Sato2024}.
However, these studies have focused on changes in the overall distribution of topological features, without examining the deformation of individual rings in detail.
To resolve the deformation modes associated with plastic rearrangements, it is necessary to establish a link between topological changes and real-space atomic displacements.

In this study, we present a topological characterisation of structural deformations during plastic rearrangements in silica glass.
We classify rearrangement events under athermal quasistatic shear according to changes in atomic coordination, and compare the topological information before and after each event.
We relate variations in topological space to real-space ring structures to examine the microscopic structural changes induced by plastic deformation.
We show that PH distinguishes intra-tetrahedral deformation from network deformation and quantifies both as continuous quantities within a unified framework.
We further find that, relative to network deformation, intra-tetrahedral deformation is more pronounced in rearrangements involving changes in coordination shells than in those preserving them.

\section{Methods}
\subsection{Model and simulations}
We perform molecular dynamics simulations of silica glass using the LAMMPS package~\cite{Thompson2022}.
The interactions are modelled using a modified form of the van Beest--Kramer--van Santen (BKS) potential~\cite{BKS1990},
\begin{align}
u_{\mu\nu}(r) = \frac{q_\mu q_\nu e^2}{4\pi\varepsilon_0 r} + A_{\mu\nu} \exp(-B_{\mu\nu} r) - \frac{C_{\mu\nu}}{r^6} + \frac{D_{\mu\nu}}{r^{24}},
\end{align}
where $r$ is the distance between two atoms of species $\mu$ and $\nu$ ($\mu, \nu \in \{ \ce{Si}, \ce{O} \}$), $q_\mu$ is the effective charge of species $\mu$ in units of the elementary charge $e$, and $\varepsilon_0$ is the vacuum permittivity.
For the parameters $q_\mu$, $A_{\mu\nu}$, $B_{\mu\nu}$, $C_{\mu\nu}$, and $D_{\mu\nu}$, we adopt the parametrisation proposed by Sundararaman \textit{et al.}; see Table~I in Ref.~\onlinecite{Sundararaman2018} for the specific values.

The potential is smoothly truncated as follows.
The Coulomb term is treated with the Wolf truncation~\cite{Wolf1999,Carre2007}
\begin{align}
 u_{\mu\nu}^W(r) = \frac{q_\mu q_\nu e^2}{4\pi\varepsilon_0} \left( \frac{1}{r} - \frac{1}{r_c^W} + \frac{r - r_c^W}{\left(r_c^W\right)^2} \right).
\end{align}
In addition, each term of the potential is multiplied by the smoothing function
\begin{align}
 G_{r_c}(r) = \exp\left( -\frac{\gamma^2}{(r - r_c)^2} \right),
\end{align}
and is set to zero for $r \geq r_c$, where $\gamma = \qty{0.2}{\angstrom}$.
This ensures that both the potential and the force vanish smoothly at the cutoff.
The cutoff length depends on the term: $r_c = r_c^W = \qty{10}{\angstrom}$ for the Coulomb term and $r_c = \qty{8}{\angstrom}$ for the other terms.

Our system consists of $N = 3000$ atoms ($N_\text{\ce{Si}} = 1000$ and $N_\text{\ce{O}} = 2000$) in a cubic box with periodic boundary conditions.
The box length is set to \qty{35.66}{\angstrom}, which corresponds to the mass density of silica glass at ambient conditions, \qty{2.2}{\gram\per\cm\cubed}~\cite{Mazurin1983}.

Starting from a random configuration, we equilibrate the system in the NVT ensemble at \qty{3500}{\kelvin} for \qty{100}{\pico\second}.
The system is then cooled to \qty{300}{\kelvin} at a constant rate of \qty{1}{\kelvin\per\pico\second} and subsequently held at \qty{300}{\kelvin} for \qty{100}{\pico\second} in the NVT ensemble.
Throughout this process, we use the Nos\'e--Hoover thermostat~\cite{Nose1984,Hoover1985} and a time step of \qty{1.2}{\femto\second}.
We repeat this procedure from different random initial configurations to obtain 300 independent samples.
Finally, we minimise the potential energy of each sample using the FIRE algorithm~\cite{Bitzek2006,Guenole2020} to obtain its inherent structure.
The termination condition for the algorithm is $F_\mathrm{max} < \qty{e-12}{\eV\per\angstrom}$.

\subsection{Athermal quasistatic shear protocol}
We shear the inherent structures using the athermal quasistatic shear (AQS) protocol~\cite{MaloneyLemaitre2006}.
Each AQS step consists of a small affine shear deformation followed by energy minimisation.
In the deformation, the $xy$ tilt factor of the box is increased by a strain increment $\delta\gamma = \num{e-5}$, defined with respect to the initial box length, and the atomic coordinates are remapped affinely.
We repeat this step 15000 times, up to a total strain of 0.15.
As the strain increases, the shear stress increases and is intermittently released through plastic rearrangements~\cite{MaloneyLemaitre2006}.
We identify a rearrangement event as a step at which the shear stress $\sigma_{xy}$ decreases from the previous step.
In total, 1693 rearrangement events are obtained from the 300 AQS runs.

\subsection{Persistent homology analysis}

\begin{figure*}
\centering
\includegraphics[width=\textwidth]{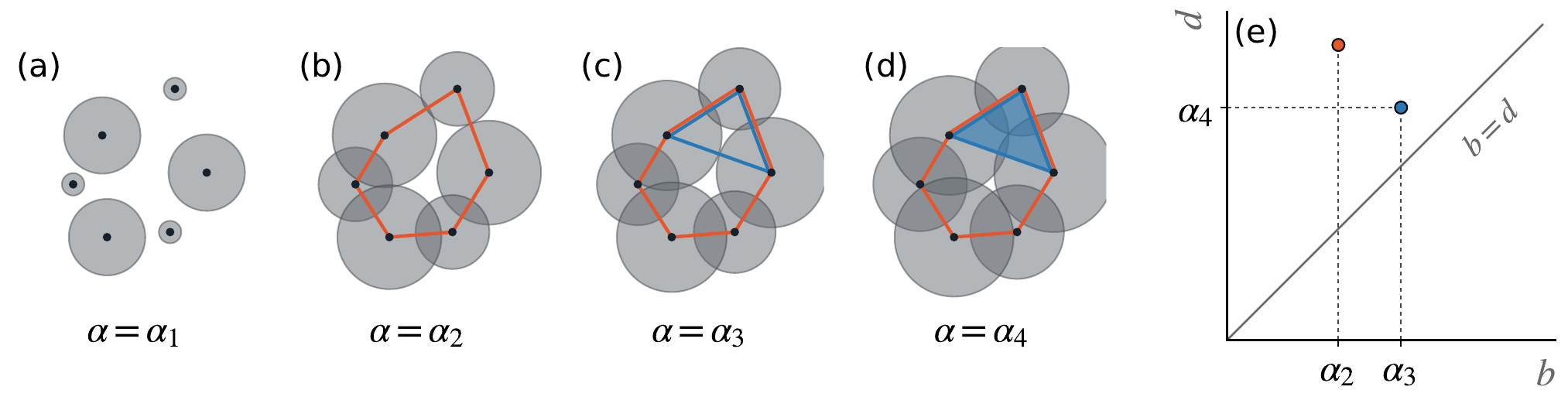}
\caption{Filtration process in persistent homology and construction of the persistence diagram.
In the filtration process, the radius of each particle is enlarged by increasing $\alpha$ (panel (a)), and eventually contacts are formed (panel (b)).
The value of $\alpha$ at which the contacting balls start to form a ring is called the birth time $b$.
As $\alpha$ increases further, smaller rings form inside the existing ring, and their birth times are defined in the same manner (panel (c)).
Finally, the voids of rings are filled (panel (d)), and the value of $\alpha$ at this stage is called the death time $d$.
The scatter plot of these birth--death time pairs is the persistence diagram (panel (e)).}
\label{fig:PH}
\end{figure*}

We then extract configurations before and after the stress drops to perform structural analysis with the PH.
In this study, we focus on rings formed by atoms, which correspond to the 1D homology.
In PH analysis (Fig.~\ref{fig:PH}), a ball is placed at each atomic position.
The radius of the ball is given by $r(\alpha)=\sqrt{\alpha + r_0^2}$,
where $\alpha$ is a parameter that controls the radius and $r_0$ is the radius at $\alpha=0$.
Following a previous study~\cite{Hiraoka2016}, we set $r_0 = \qty{0.375}{\angstrom}$ for \ce{Si} and \qty{1.275}{\angstrom} for \ce{O}.
As $\alpha$ increases (Figs.~\ref{fig:PH}(a)--\ref{fig:PH}(d)), the balls grow and begin to overlap, and edges and triangles are successively added according to the overlaps.
The resulting sequence of structures is called a filtration~\cite{Edelsbrunner2000,Zomorodian2005}.
At a certain value of $\alpha$, a ring is formed by the overlapping balls;
for example, a ring consisting of six balls appears in Fig.~\ref{fig:PH}(b),
and rings consisting of three balls appear in Fig.~\ref{fig:PH}(c).
This value of $\alpha$ is called the birth time $b$.
As $\alpha$ increases further, the ring is eventually filled in and disappears;
for example, the ring consisting of three balls disappears in Fig.~\ref{fig:PH}(d).
This value of $\alpha$ is called the death time $d$, and the difference $l = d - b$ is called the lifetime.
The scatter plot of the birth and death times of all rings, as shown in Fig.~\ref{fig:PH}(e), is called the persistence diagram (PD)~\cite{Edelsbrunner2000,Obayashi2022}.

PH analysis is performed using the HomCloud package~\cite{Obayashi2022,HomCloud}, in which the filtration is constructed from the weighted alpha complex~\cite{Edelsbrunner1995,Hiraoka2016}.
To take into account the periodic boundary conditions of the parallelepiped cell, a $3 \times 3 \times 3$ supercell is constructed, and the central cell together with its surrounding margin region is extracted from the supercell.
The width of the margin is set to \qty{10}{\percent} of the cell length in each direction.

To identify the ring corresponding to each point in a PD, we use the optimal volume~\cite{Obayashi2018} and the stable volume~\cite{Obayashi2023} implemented in the HomCloud.
Specifically, stable volumes with a noise level of \qty{0.001}{\square\angstrom} are used for PD points with $l > \qty{0.1}{\square\angstrom}$, and optimal volumes are used for the other PD points~\cite{Minamitani2025}.
This procedure gives the set of atoms that constitutes each ring.
We consider only rings that contain at least one atom in the central cell, and replicated atoms are identified with their original atoms.
Because of the margin region, rings consisting of the same set of atoms may be obtained more than once;
such rings are counted only once.
Hereafter, the PDs are constructed only from the rings obtained in this manner.

\section{Two types of plastic rearrangements}
\label{sec:two_types}
First, we classify the rearrangement events into two types according to whether the coordination environment of \ce{Si} atoms changes~\cite{Koziatek2015,Bonfanti2019}.
An \ce{O} atom is regarded as coordinating a \ce{Si} atom if their separation is smaller than a threshold of \qty{2.175}{\angstrom}, which corresponds to the first minimum of the \ce{Si}--\ce{O} pair correlation function at zero strain.
This value is kept fixed throughout the deformation.
In silica glass, each \ce{Si} atom is predominantly coordinated by four \ce{O} atoms to form a \ce{SiO4} tetrahedron, and the tetrahedra form a three-dimensional network by sharing their corner \ce{O} atoms~\cite{Zachariasen1932,Elliott1991}.
Indeed, in our configurations, \qty{99.95}{\percent} of the \ce{Si} atoms are fourfold coordinated at zero strain, so that nearly every \ce{Si} atom forms a \ce{SiO4} tetrahedron with its coordinating \ce{O} atoms.

We refer to an event in which the indices of coordinating \ce{O} atoms remain unchanged for all \ce{Si} atoms as a coordination-preserving (CP) event, and to an event in which the indices change for at least one \ce{Si} atom as a coordination-changing (CC) event.
CC events thus include not only changes in the coordination number but also exchanges of coordinating \ce{O} atoms that leave the coordination number unchanged.
Of the 1693 events obtained in the present study, 1271 are CP events and 422 are CC events.
Unless otherwise noted, the statistical analyses below are performed on all these events.

\begin{figure}
\centering
\includegraphics[width=\columnwidth]{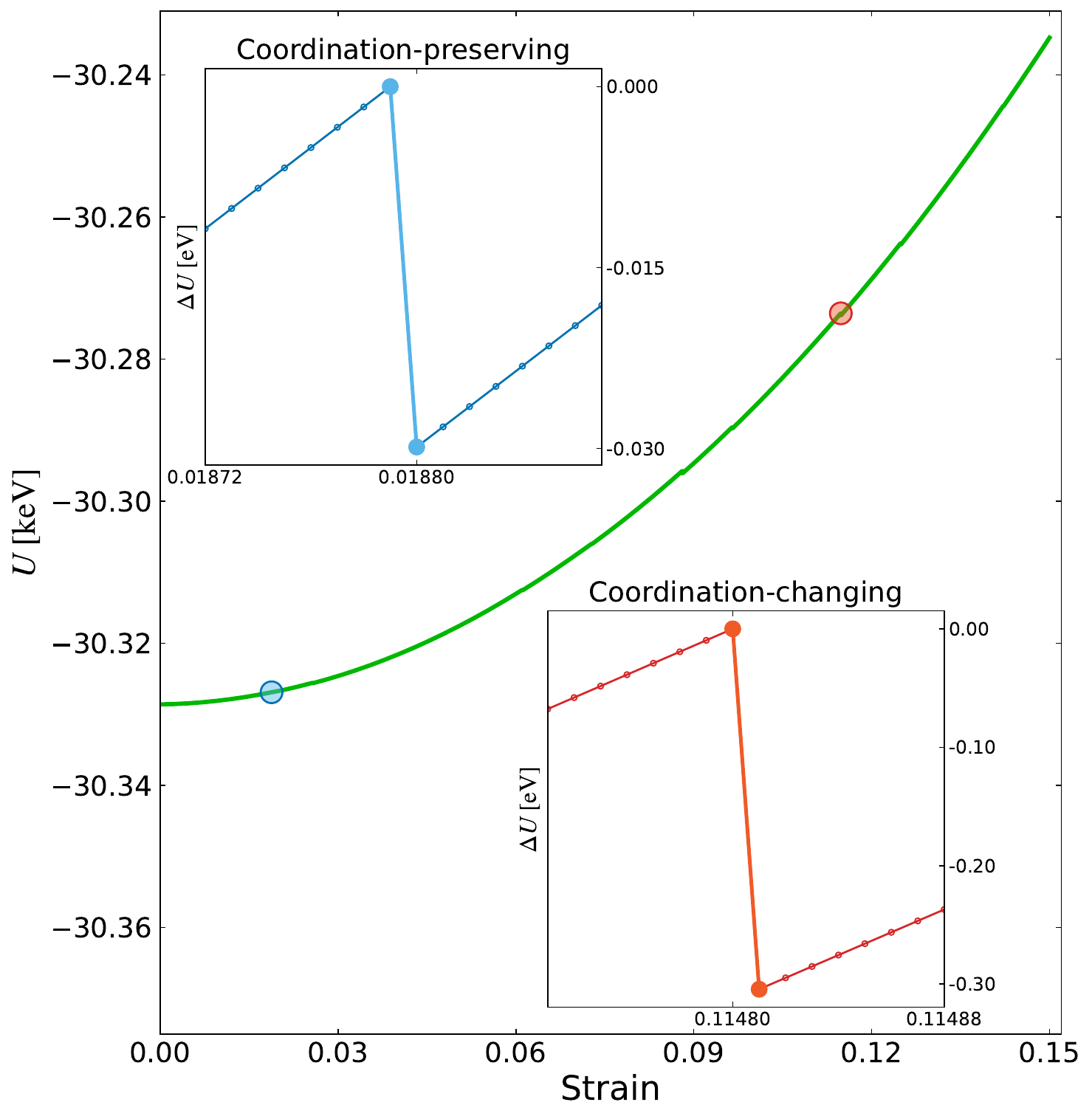}
\caption{The potential energy $U$ in a representative AQS run.
The insets show magnified views of the two rearrangement events marked by circles on the main curve.
In each inset, the origin of $y$-axis is shifted to the value of energy before the drop.}
\label{fig:energy-strain}
\end{figure}

Figure~\ref{fig:energy-strain} shows the potential energy during the AQS run of a representative sample.
The average energy drop is \qty{0.0357}{\eV} for CP events and \qty{0.377}{\eV} for CC events; the energy drops in CC events are about ten times larger.
Therefore, the structural classification of bond coordination is able to distinguish two types of events with different energetic characteristics.
However, because this classification is based on the binary judgement of \ce{Si}--\ce{O} connectivity, it quantifies neither the magnitude of the structural change nor which structural motifs change upon rearrangement.

\section{Change in the persistence diagram}
\begin{figure}
\centering
\includegraphics[width=\columnwidth]{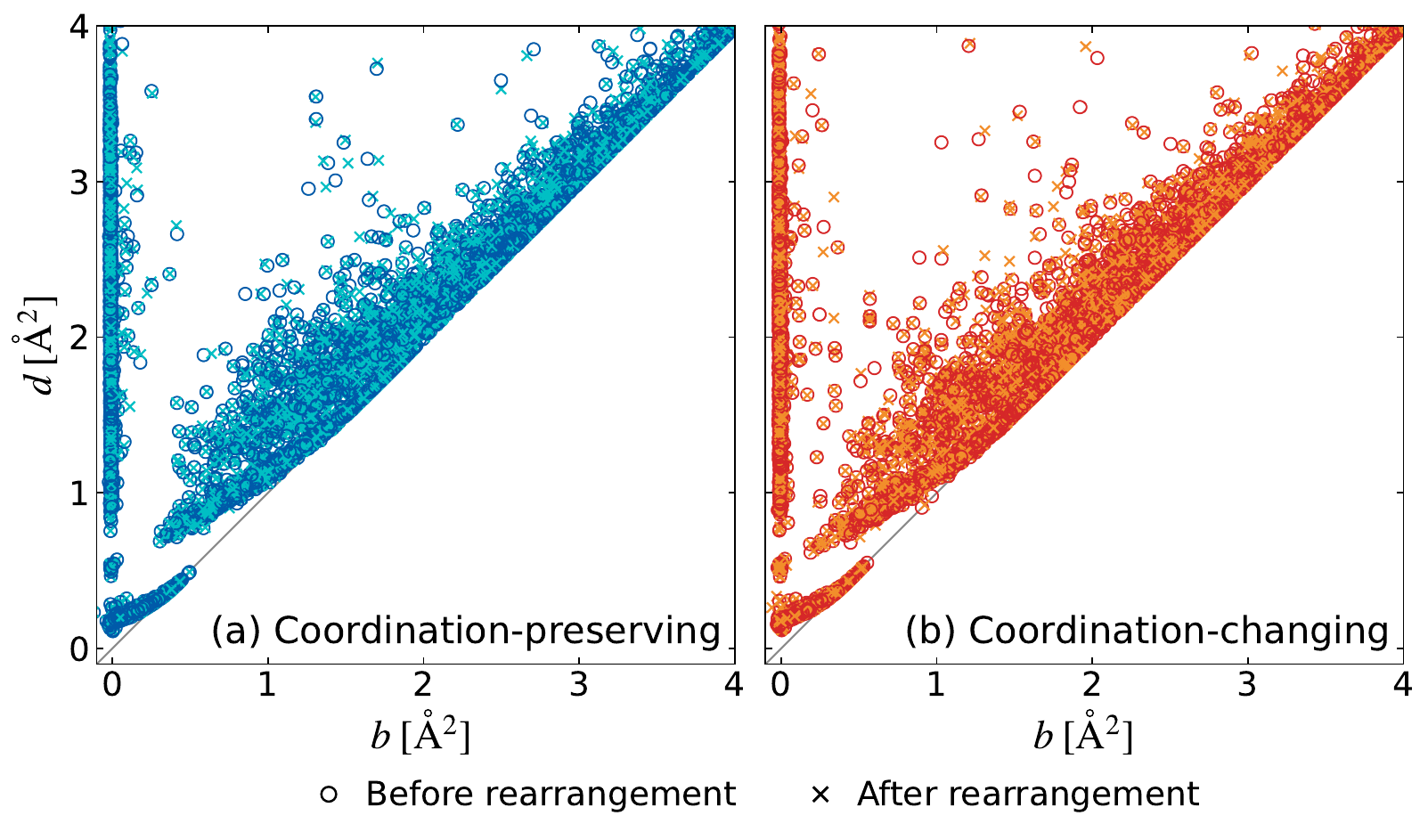}
\caption{Persistence diagrams before and after the two rearrangement events marked in Fig.~\ref{fig:energy-strain}:
Panels (a) and (b) show coordination-preserving and coordination-changing events, respectively.
Circles and crosses represent the birth--death pairs in the configurations before and after the rearrangement, respectively.}
\label{fig:pd_before_after}
\end{figure}

To address these points, we first evaluate the overall magnitude of the structural change using PH.
We compare the PDs of the configurations before and after the two energy drops marked in Fig.~\ref{fig:energy-strain}.
As shown in Fig.~\ref{fig:pd_before_after}, the birth--death pairs exhibit more pronounced shifts in the CC event (panel (b)) than in the CP event (panel (a)).
To quantify this difference, we compute the 2-Wasserstein distance $W_2$~\cite{CohenSteiner2010} between the two PDs.
The Wasserstein distance is defined as the minimum cost over all matchings between the points of two PDs, where points may also be matched to the diagonal, and thus quantifies the overall dissimilarity between them~\cite{Aktas2019}.
For two PDs $D$ and $D'$, the $p$-Wasserstein distance is defined as
\begin{align}
W_p(D, D') = \left( \inf_{\eta} \sum_{x \in D \cup \Delta} \| x - \eta(x) \|_2^p \right)^{1/p},
\end{align}
where $\Delta$ denotes the diagonal $\{(b, d) \mid b = d\}$, whose points are counted with infinite multiplicity, $\eta$ ranges over all bijections from $D \cup \Delta$ to $D' \cup \Delta$, and $\| \cdot \|_2$ denotes the Euclidean norm in the birth--death plane.
We use $p = 2$.
The Wasserstein distance between PDs is computed using the Hera implementation~\cite{Kerber2017} in GUDHI~\cite{gudhi:WassersteinDistance}, with the relative error parameter set to $\delta = 0.01$.
Averaged over all events, we obtain $W_2 = \qty{1.41 \pm 0.44}{\angstrom\squared}$ for CP events and $W_2 = \qty{1.62 \pm 0.48}{\angstrom\squared}$ for CC events, where the errors denote standard deviations.
Thus, although most birth--death pairs overlap before and after the rearrangement in both types of events, the overall structural change is larger on average in CC events.

\section{Real-space tracking of topological changes}
\label{sec:realspace}
To connect the changes observed in the PD and real-space rearrangements, we perform the inverse analysis.
This analysis associates each point in the PD with a ring of atoms~\cite{Hiraoka2016,Nakamura2015,Onodera2020}.
By this analysis, we obtain the ring vertex indices of each point in the PD and reveal that the majority of rings do not change their vertex atoms through the rearrangements.
We refer to such rings as ``surviving rings'' and to the other rings in the configuration before the rearrangement as ``vanishing rings.''
The average fractions of vanishing rings are \qty{2.67}{\percent} for CP events and \qty{3.78}{\percent} for CC events; that is, rings disappear more frequently in CC events.
Rings that appear only in the configuration after the rearrangement fall into neither category and are not considered in this study.

\begin{figure}
\centering
\includegraphics[width=\columnwidth]{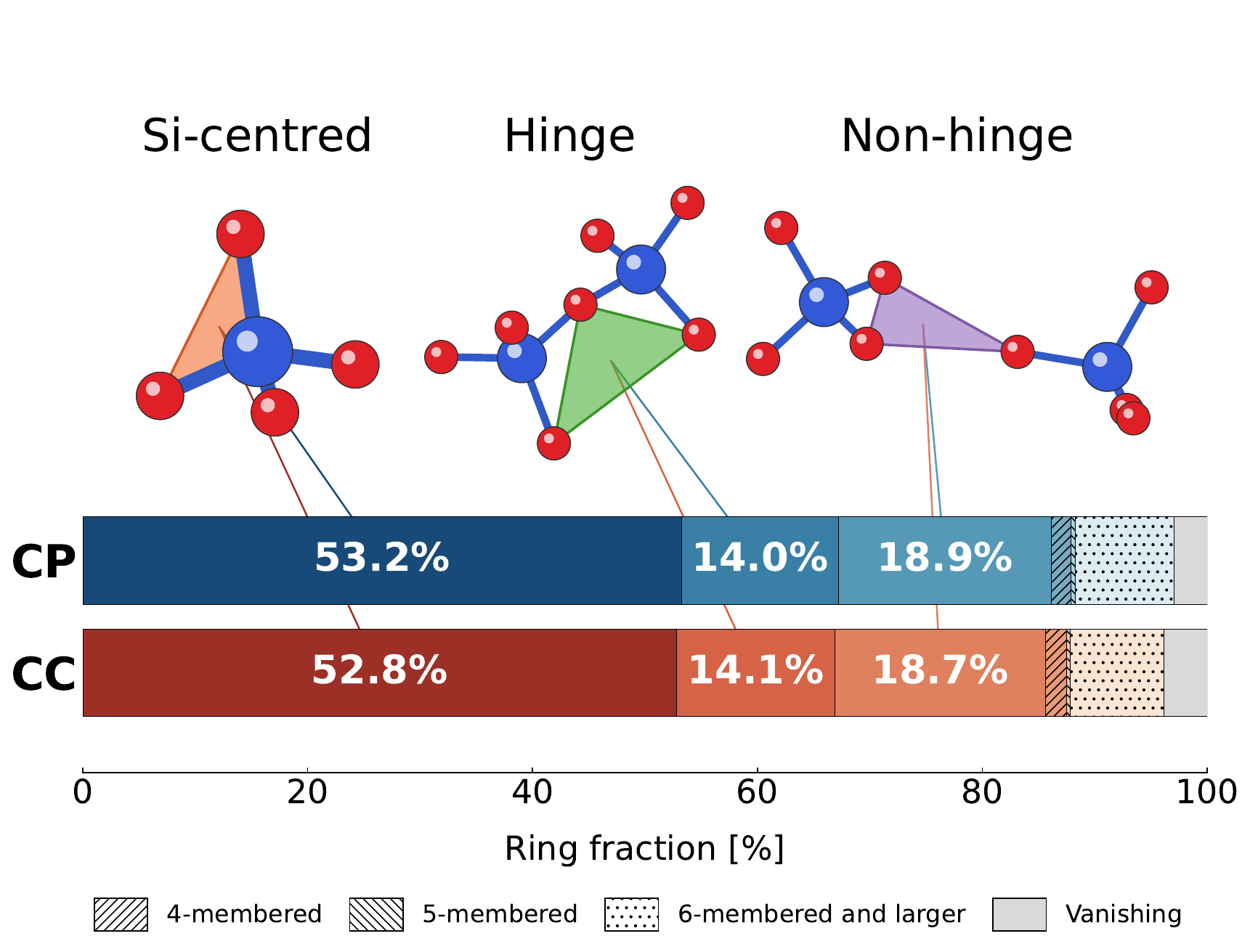}
\caption{Size composition of the rings in the configurations before the two rearrangements shown in Fig.~\ref{fig:energy-strain}.
The coloured and hatched segments represent surviving rings, and the grey segments represent vanishing rings, i.e., rings whose vertex atoms change upon the rearrangement.
The schematics illustrate the three types of triangles, with blue and red spheres representing \ce{Si} and \ce{O} atoms, respectively.}
\label{fig:ring_size_composition}
\end{figure}

Composition of surviving rings is shown in Fig.~\ref{fig:ring_size_composition} for the two types of rearrangement events.
In both events, the surviving rings are dominated by three-membered rings, i.e., triangles.
These triangles can be further classified into three types, as illustrated by the schematics in the figure.
The first type is the \ce{Si1O2} triangle (referred to as the \ce{Si}-centred triangle), which consists of one \ce{Si} atom and two \ce{O} atoms, and is part of a \ce{SiO4} tetrahedron~\cite{Hiraoka2016}.
The other two types are \ce{O3} triangles, which consist of three \ce{O} atoms; we further classify them according to the number of their triangular edges that also form \ce{O}--\ce{O} edges of \ce{SiO4} tetrahedra.
We call an \ce{O3} triangle a ``hinge triangle'' if two or three of its edges are \ce{O}--\ce{O} edges of different tetrahedra, because such a triangle is formed between neighbouring tetrahedra sharing a corner \ce{O} atom.
We call the other \ce{O3} triangles ``non-hinge triangles''.
Of all surviving hinge triangles found in our samples, only \qty{0.17}{\percent} have three such edges; hinge triangles are thus almost exclusively formed by two edges belonging to two tetrahedra that share a corner \ce{O} atom.
Moreover, among the hinge triangles with exactly two such edges, only \qty{0.0019}{\percent} of the triangles have non-tetrahedral edge shorter than the two tetrahedral \ce{O}--\ce{O} edges.
That is, the longest triangular edge is a non-tetrahedral edge in the vast majority of hinge triangles.
On the other hand, of all surviving non-hinge triangles, \qty{96.32}{\percent} have exactly one tetrahedral \ce{O}--\ce{O} edge.
\qty{3.67}{\percent} of them do not contain any tetrahedral edge; all three \ce{O} atoms belong to different tetrahedra.
Finally, the remaining \qty{0.01}{\percent} of the non-hinge triangles are comprised of a face of \ce{SiO4} tetrahedra or \ce{O} atoms that do not have any connection with fourfold coordinated \ce{Si} atoms.
In the overall composition (Fig.~\ref{fig:ring_size_composition}), rings with four or more vertices account for less than \qty{15}{\percent} of the surviving rings, and we therefore focus on triangles in what follows.

\begin{figure}
\centering
\includegraphics[width=\columnwidth]{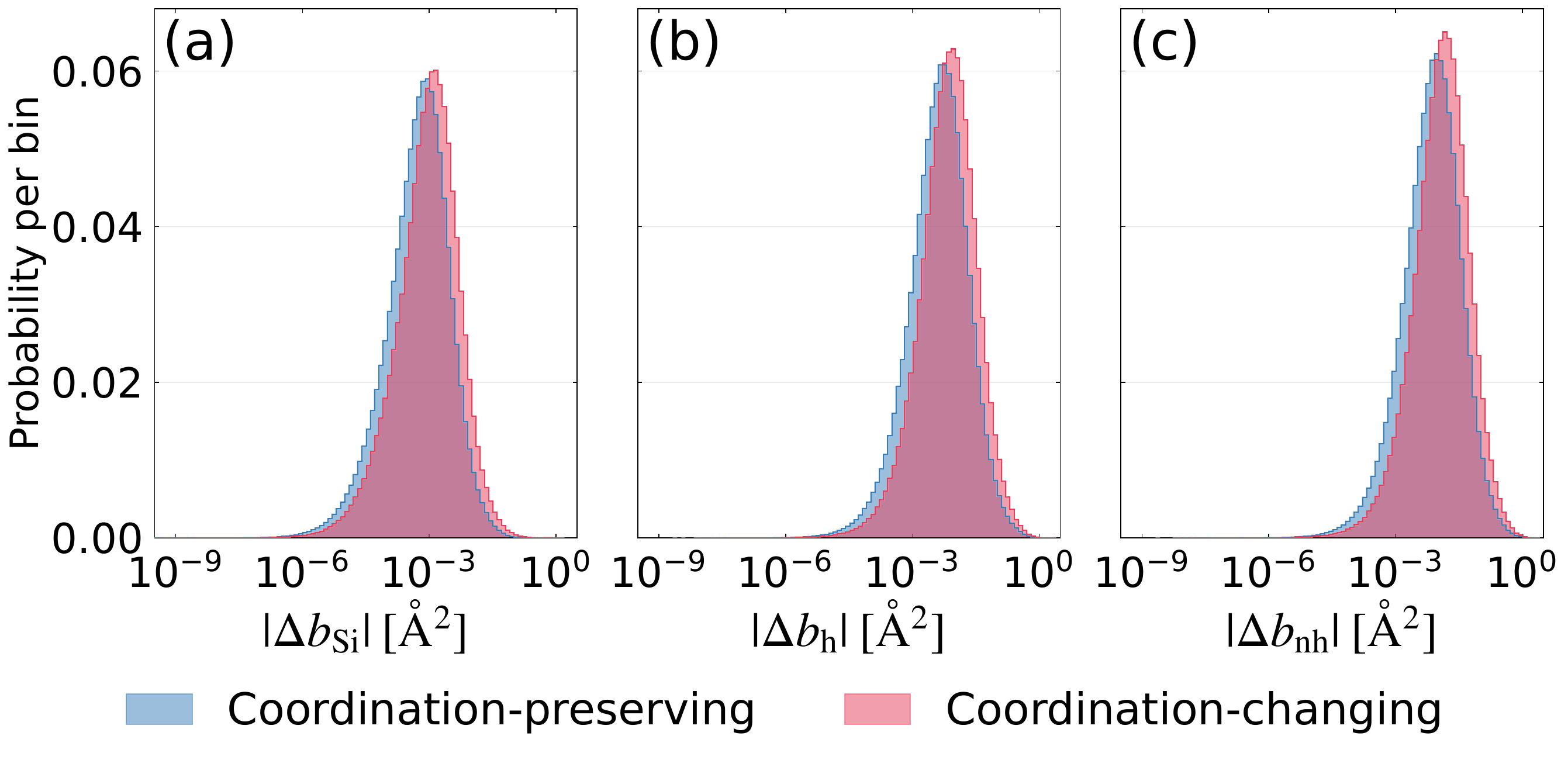}
\caption{Distributions of the absolute change in birth time $\lvert \Delta b \rvert$ in the two types of rearrangement events for (a) \ce{Si}-centred, (b) hinge, and (c) non-hinge triangles.}
\label{fig:birth_time_change_distributions}
\end{figure}

Of the quantities that PH assigns to each ring, namely the birth time $b$, the death time $d$, and the lifetime $l$, we focus on the birth time, which for a triangle is directly related to its edge lengths, as discussed below.
For each surviving ring, we define the change in birth time as $\Delta b = b_\text{after} - b_\text{before}$, and denote it by $\Delta b_\mathrm{Si}$, $\Delta b_\mathrm{h}$, and $\Delta b_\mathrm{nh}$ for \ce{Si}-centred, hinge, and non-hinge triangles, respectively.
Figure~\ref{fig:birth_time_change_distributions} shows the distributions of $\lvert \Delta b \rvert$ over all rearrangement events.
For all three types of triangles, the distributions for CC events extend to larger values of $\lvert \Delta b \rvert$ than those for CP events, indicating that the PD points shift more in CC events.

To clarify the meaning of $\Delta b$, we relate it to the geometry of the triangles.
In the alpha filtration, a triangle forms a ring when the last of its three edges enters the filtration, provided that this edge does not enter simultaneously with the triangle itself.
For every triangle considered here, this last edge is the longest edge, which is an \ce{O}--\ce{O} edge; because its two end atoms have the same radius, $\Delta b = \Delta(\ell^2)/4$, where $\ell$ is its length.
If $a$ and $c$ denote the lengths of the two edges adjacent to the longest edge and $\theta$ the angle between them, the law of cosines gives
\begin{align}
 \Delta b \simeq \frac{ac}{2} \Delta\theta \sin\theta + \frac{\Delta (a^2 + c^2) - 2 \Delta(ac) \cos\theta}{4}
 \label{eq:delta_b_expansion}
\end{align}
for small changes.
When $a$ and $c$ are nearly constant, the second term is negligible and $\Delta b$ is proportional to $\Delta\theta$.

In an \ce{Si}-centred triangle, $a = c = r_\mathrm{SiO}$ is the nearly constant \ce{Si}--\ce{O} bond length and $\theta$ is the \ce{O}--\ce{Si}--\ce{O} angle, so that $\Delta b_\mathrm{Si} \simeq \frac{1}{2} r_\mathrm{SiO}^2 \Delta\theta \sin\theta$.
A similar argument applies to hinge triangles.
Since all three vertices are \ce{O} atoms, a hinge triangle can form a ring only if the angle $\varphi$ at the shared \ce{O} atom is smaller than \ang{90};
otherwise, the shared \ce{O} atom lies inside the sphere having the opposite edge as its diameter, so that this edge enters the filtration simultaneously with the triangle and no ring is born.
For $\varphi > \ang{60}$, the longest edge is the one connecting the two non-shared \ce{O} atoms, so that $\theta = \varphi$ and $a$ and $c$ are the nearly rigid tetrahedral \ce{O}--\ce{O} edges; $\Delta b_\mathrm{h}$ thus reflects the change in $\varphi$, that is, the opening or closing of the hinge.
For non-hinge triangles, in contrast, at least one of the two edges adjacent to the longest edge connects \ce{O} atoms that do not belong to a common tetrahedron.
Its length is not constrained by the rigid tetrahedra, so that the second term in Eq.~\eqref{eq:delta_b_expansion} is not negligible and $\Delta b_\mathrm{nh}$ has no simple interpretation in terms of a single angle.

\begin{figure}
\centering
\includegraphics[width=\columnwidth]{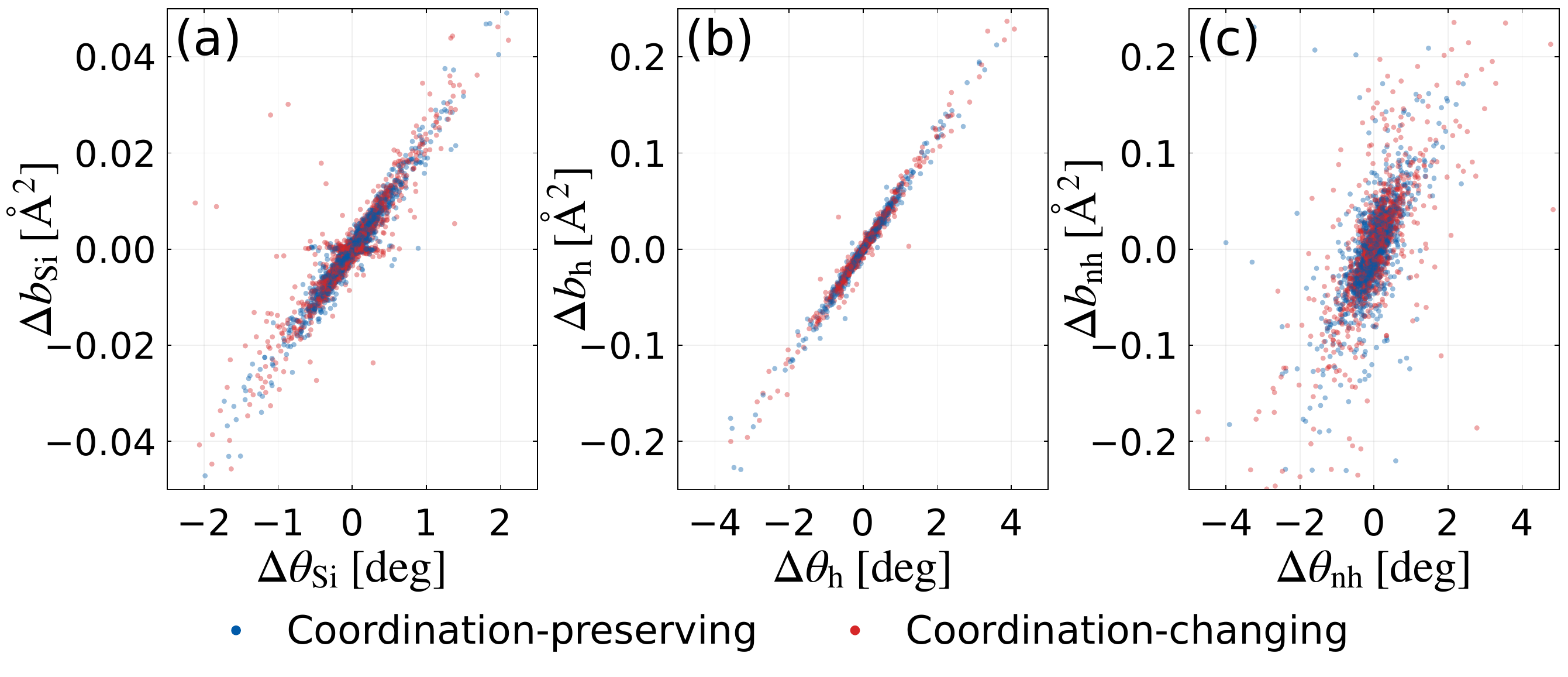}
\caption{Change in birth time $\Delta b$ versus the change in the angle opposite the longest edge $\Delta \theta$ for individual surviving triangles in the two rearrangement events shown in Fig.~\ref{fig:energy-strain}:
(a) \ce{Si}-centred, (b) hinge, and (c) non-hinge triangles.
Colours distinguish the CP and CC events.}
\label{fig:birth_angle_correspondence}
\end{figure}

To test these expectations, we define, for each surviving triangle, $\theta$ as the angle opposite its longest edge in the configuration before the rearrangement and $\Delta \theta = \theta_\text{after} - \theta_\text{before}$ as the change in the angle at the same vertex.
This vertex is kept fixed even if a different edge becomes the longest after the rearrangement.
We denote $\theta$ for \ce{Si}-centred, hinge, and non-hinge triangles by $\theta_\text{\ce{Si}}$, $\theta_\mathrm{h}$, and $\theta_\mathrm{nh}$, respectively;
$\theta_\text{\ce{Si}}$ is an \ce{O}--\ce{Si}--\ce{O} angle, whereas $\theta_\mathrm{h}$ and $\theta_\mathrm{nh}$ are \ce{O}--\ce{O}--\ce{O} angles.
Note that $\theta$ is evaluated only for the triangles identified by the inverse analysis, not for all triplets of atoms.

Figure~\ref{fig:birth_angle_correspondence} plots $\Delta b$ against $\Delta \theta$ for individual surviving triangles in the two events presented in Fig.~\ref{fig:energy-strain}.
For \ce{Si}-centred triangles, the data fall nearly on a single straight line (panel (a)), confirming that $\Delta b_\mathrm{Si}$ essentially measures the change in the \ce{O}--\ce{Si}--\ce{O} angle.
Hinge triangles show a similarly tight linear relation (panel (b)).
As shown above, in almost all surviving hinge triangles, the longest edge connects the two non-shared \ce{O} atoms, so that $a$ and $c$ in Eq.~\eqref{eq:delta_b_expansion} are the rigid tetrahedral \ce{O}--\ce{O} edges and $\Delta b_\mathrm{h}$ essentially measures the opening of the hinge.
For non-hinge triangles, by contrast, $\Delta \theta_\mathrm{nh}$ and $\Delta b_\mathrm{nh}$ are positively correlated but widely scattered (panel (c)), reflecting the non-negligible second term in Eq.~\eqref{eq:delta_b_expansion}.
For all three types, triangles in CP and CC events follow the same relations.

These results show that $\Delta b$ is not merely an abstract
topological quantity but directly reflects the deformation of the atomic network in real space.
For \ce{Si}-centred and hinge triangles, $\Delta b$ corresponds to physically intuitive quantities, namely the \ce{O}--\ce{Si}--\ce{O} angle and the opening of the hinge between corner-sharing tetrahedra, respectively.
For non-hinge triangles, $\Delta b$ captures combined changes in angles and inter-tetrahedral distances that lack a simple geometric description;
$\Delta b$ thus provides a single, continuous measure applicable both to deformations with and without a simple geometric interpretation.

\begin{figure}
\centering
\includegraphics[width=\columnwidth]{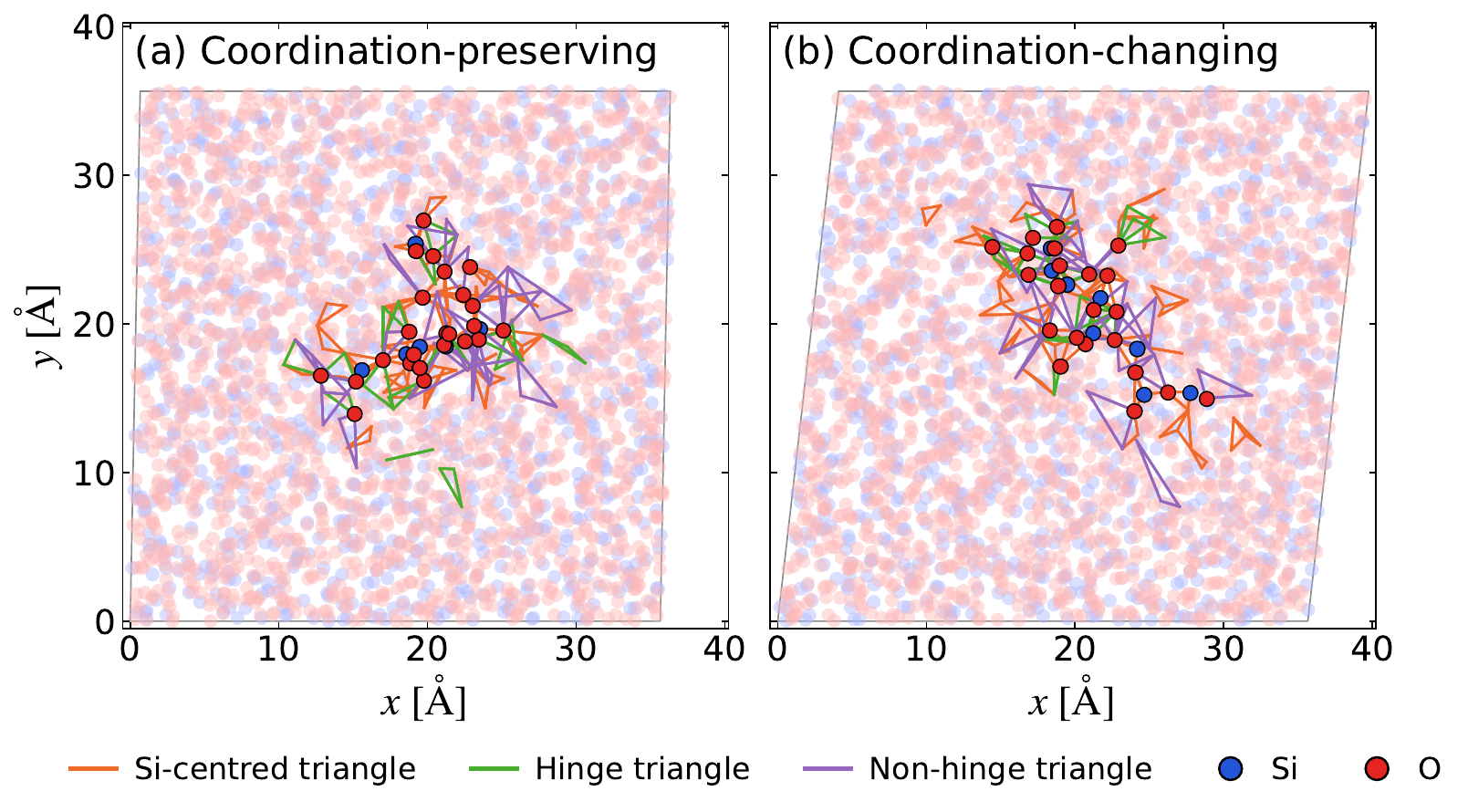}
\caption{Projections onto the $xy$ plane of atoms with large displacements and triangles with large $\lvert \Delta b \rvert$ in the two rearrangement events shown in Fig.~\ref{fig:energy-strain}: (a) the CP event and (b) the CC event.
The atoms shown are the \qty{1}{\percent} of all atoms (\ce{Si} and \ce{O} combined) with the largest displacements upon the rearrangement, and the triangles shown are, for each type, the \qty{1}{\percent} of the surviving triangles with the largest $\lvert \Delta b \rvert$.
Pale circles represent all atoms in the system.
Panel (a) shows 6 \ce{Si} atoms, 24 \ce{O} atoms, 60 \ce{Si}-centred triangles, 16 hinge triangles, and 22 non-hinge triangles, and panel (b) shows 8 \ce{Si} atoms, 22 \ce{O} atoms, 60 \ce{Si}-centred triangles, 16 hinge triangles, and 21 non-hinge triangles.}
\label{fig:md_pd_spatial_projection}
\end{figure}

Finally, we examine where the triangles with large $\lvert \Delta b \rvert$ are located.
If $\Delta b$ reflects the local deformation of triangles, such triangles should be concentrated in the rearrangement region identified from atomic displacements.
Figure~\ref{fig:md_pd_spatial_projection} compares the two.
In both the CP and CC events (panels (a) and (b), respectively), the triangles with large $\lvert \Delta b \rvert$ are located where the atoms with large displacements are concentrated.
Thus, triangles with large $\lvert \Delta b \rvert$ are localised in the region of the plastic rearrangement, and $\Delta b$ captures the local structural changes associated with the rearrangement.

So far, we have shown that $\Delta b$ directly reflects the real-space deformation of triangles in the rearrangement region.
However, CC events differ from CP events in overall magnitude, as shown by their larger energy drops (Sec.~\ref{sec:two_types}) and larger $W_2$.
The shifts of the distributions in Fig.~\ref{fig:birth_time_change_distributions} may therefore simply reflect this difference in magnitude rather than a difference in the character of the structural changes.
To compare the character of the two types of events, we need quantities that are independent of the event magnitude.

\section{Deformation modes in two types of rearrangements}
As described above, \ce{Si}-centred triangles are formed inside \ce{SiO4} tetrahedra, whereas hinge and non-hinge triangles are formed in the network connecting the tetrahedra.
The behaviour of \ce{Si}-centred triangles is therefore expected to reflect intra-tetrahedral deformation~\cite{Pedone2008}, and that of hinge and non-hinge triangles the deformation of the tetrahedral network.
Of the latter two, hinge triangles have a clearer geometric meaning:
$\Delta b_\mathrm{h}$ essentially measures the opening of the hinge between corner-sharing tetrahedra (Fig.~\ref{fig:birth_angle_correspondence}(b)), whereas $\Delta b_\mathrm{nh}$ mixes changes in angles and in inter-tetrahedral \ce{O}--\ce{O} distances.
We therefore use \ce{Si}-centred and hinge triangles to compare the relative contributions of the two types of deformation in CP and CC events.

For each rearrangement event $e$ and each type of triangle $X \in \{\text{\ce{Si}}, \mathrm{h}\}$, we calculate the average of $\lvert \Delta b \rvert$ in the event and denote it by $\langle \lvert \Delta b_{X} \rvert \rangle_e$.
We then define the ratio
\begin{align}
 R_b^{(e)} = \frac{\langle \lvert \Delta b_\text{\ce{Si}} \rvert \rangle_e}{\langle \lvert \Delta b_\mathrm{h} \rvert \rangle_e},
 \label{eq:ratio_b}
\end{align}
which measures the magnitude of intra-tetrahedral deformation $\Delta b_\text{\ce{Si}}$ relative to that of network deformation $\Delta b_\mathrm{h}$ in event $e$.
Because both the numerator and the denominator scale with the overall magnitude of the rearrangement, taking the ratio largely cancels out differences in event magnitude.
Note that the absolute value of $R_b^{(e)}$ does not indicate which type of deformation dominates, because $\Delta b$ responds to a given change in angle with a different sensitivity for each type of triangle (Sec.~\ref{sec:realspace}); only comparisons of $R_b^{(e)}$ between events are meaningful.
Note also that in CC events, \ce{Si}-centred triangles whose vertices change upon rearrangement are not surviving triangles and thus are not included in the average; this exclusion, if anything, lowers $R_b^{(e)}$ for CC events.

\begin{figure}
\centering
\includegraphics[width=\columnwidth]{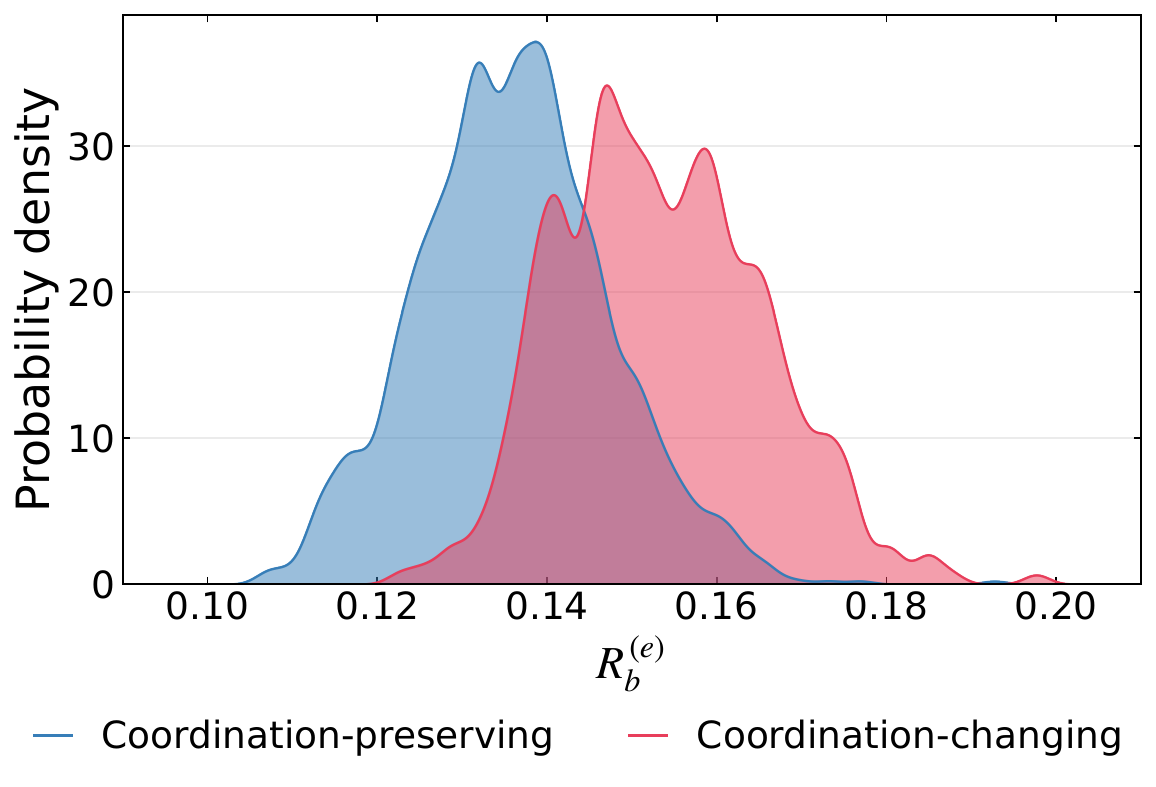}
\caption{Distributions of the ratio $R_b^{(e)}$ (Eq.~\eqref{eq:ratio_b}) for the two types of rearrangement events.
The distributions are kernel density estimates with a Gaussian kernel of bandwidth $0.0015$.
The distribution for CC events is shifted to larger values, indicating that intra-tetrahedral deformation is relatively larger in CC events.}
\label{fig:event_birth_angle_ratios}
\end{figure}

Figure~\ref{fig:event_birth_angle_ratios} shows the distributions of $R_b^{(e)}$ for the two types of rearrangement events.
The distribution for CC events is shifted toward larger values than that for CP events:
$R_b^{(e)} = \num{0.136 \pm 0.011}$ for CP events and \num{0.153 \pm 0.012} for CC events (mean $\pm$ standard deviation).
This indicates that, relative to network deformation, intra-tetrahedral deformation is larger in rearrangements that change the coordination shells.

\section{Discussion}
In this study, we perform persistent homology analysis to extract the structural change in the plastic rearrangement of silica glass.
By inverse analysis, we identified \ce{Si}-centred triangles, hinge triangles, and non-hinge triangles, which constitute the majority of the surviving rings, and study their changes of birth times $\Delta b$.
Our results can be summarised as follows:
(i) $\Delta b_\text{\ce{Si}}$ and $\Delta b_\mathrm{h}$ are nearly proportional to the change in the angle opposite the longest edge, which is the \ce{O}--\ce{Si}--\ce{O} angle for \ce{Si}-centred triangles and the opening angle of the hinge for hinge triangles.
In contrast, $\Delta b_\mathrm{nh}$ is positively correlated with the change in this angle but widely scattered, because it also reflects changes in the inter-tetrahedral \ce{O}--\ce{O} distances.
(ii) Triangles with large $\lvert \Delta b \rvert$ are localised in the rearrangement region, where atomic displacements are large.
(iii) The ratio $R_b^{(e)}$ is larger for CC events, indicating that intra-tetrahedral deformation is larger relative to network deformation in CC events.

Result (iii) may be related to the larger energy drops in CC events (Sec.~\ref{sec:two_types}).
The \ce{O}--\ce{Si}--\ce{O} angles within a tetrahedron are much stiffer than the inter-tetrahedral degrees of freedom, such as the \ce{Si}--\ce{O}--\ce{Si} angles and the rotations of tetrahedra, which determine the shape of hinge triangles~\cite{Elliott1991,Mantisi2012}.
A given amount of intra-tetrahedral deformation therefore involves a larger change in energy than the same amount of network deformation.
We thus speculate that, in CC events, the tetrahedral strain accumulated under shear is released together with bond switching, and that this release contributes to the larger energy drops.
Note, however, that the \ce{Si}-centred triangles around a \ce{Si} atom whose coordination shell changes are necessarily distorted by the bond switching itself.
Whether the increase in $R_b^{(e)}$ reflects such deformation localised at the switching site or a more widespread release of tetrahedral strain remains to be clarified.

The vanishing rings provide complementary information.
In bond-based ring analyses, bond switching destroys existing rings and creates new ones~\cite{Nomura2011}.
Consistently, the fraction of vanishing rings is larger in CC events (\qty{3.78}{\percent}) than in CP events (\qty{2.67}{\percent}).
However, vanishing rings are also found in CP events, in which no \ce{Si}--\ce{O} bond is broken.
Because rings in PH are defined from the atomic geometry rather than from bonds, they can also disappear through geometric deformation alone.

Taken together, our results show that PH separately captures intra-tetrahedral and network deformation as continuous quantities within a single framework, which coordination analysis cannot do.
Coordination analysis indicates only whether the \ce{Si}--\ce{O} connectivity changes, whereas in PH, intra-tetrahedral deformation appears in the \ce{Si}-centred triangles and network deformation in the hinge and non-hinge triangles.
This separation is not trivial.
When \ce{O}--\ce{O}--\ce{O} angles are defined using a distance cutoff~\cite{LeRoux2010,Firooz2024}, the angles of the tetrahedral faces are also included, so that the two types of deformation are mixed.
In contrast, PH detects virtually no \ce{O3} triangles whose vertices all belong to the same tetrahedron,
and it further isolates hinge triangles, whose birth times essentially measure the opening of the hinges between corner-sharing tetrahedra.
Beyond separating these deformation modes, PH also distinguishes, within the same analysis, continuous changes in the surviving rings from the loss of rings, i.e., the vanishing rings.

Our results were obtained under the following conditions.
The system contains 3000 atoms, only a single cooling rate is considered~\cite{Vollmayr1996}, and the deformation is applied by the AQS, which excludes thermal fluctuations.
In addition, the analysis is limited to the birth times of triangles among the surviving rings; larger rings and other quantities such as the death time and lifetime are not used.
Extending the analysis to larger rings, death times, and vanishing rings is expected to capture changes in medium-range order~\cite{Minamitani2025,Shiga2024}.
Moreover, whereas this study characterised how the structure changes during a rearrangement, whether the location of a rearrangement or its type (CP or CC) can be predicted from the PD of the structure before the rearrangement remains an open question~\cite{Richard2020Predicting}.
Answering this question would connect the present analysis to the long-standing problem of identifying where plastic rearrangements occur.

\begin{acknowledgments}
This work is supported by JST FOREST Program (Grant Number JPMJFR236Q) and JSPS KAKENHI Grant Numbers 25H01478 and 26K00672.
\end{acknowledgments}

\section*{Conflict of Interest}
The authors have no conflicts to disclose.

\section*{Author contributions}
\textbf{F.~Katsuyama:} Conceptualization, Data curation, Formal analysis, Investigation, Methodology, Validation, Visualization, Writing -- original draft, Writing -- review \& editing;
\textbf{K.~Shiraishi:} Conceptualization, Formal analysis, Investigation, Methodology, Project administration, Supervision, Validation, Writing -- review \& editing;
\textbf{E.~Minamitani:} Conceptualization, Funding acquisition, Formal analysis, Investigation, Methodology, Project administration, Resources, Supervision, Validation, Writing -- review \& editing.

\end{document}